# Making sense of Physics[1] in the first year of study

Shirley Booth
Centre for educational development
Chalmers University of Technology

Åke Ingerman
Department of Physics (MiNa)
Chalmers University of Technology

## *Abstract*

We address the question "How do students make sense of Physics from the point of view of constituting physics knowledge?". A phenomenographic study is described as a result of which we present six qualitatively different ways in which students experience the first year of Physics. The variation is analysed in terms of the structure of experience, the nature of knowledge and an ethical aspect related to the identification of authority. Three of these ways of experiencing the first year are considered to be unproductive in terms of making sense of physics, while the other three support to an increasing degree the formation of a well-grounded physics knowledge object. Implications for practice are considered.

## *1. Aim*

Students beginning to study engineering physics are faced in their first year with a bewildering array of new subjects and teachers. Such is the case at Chalmers University of Technology in Göteborg, Sweden, where our study took place. At the heart of the four and a half year programme is a vision of the all-round engineer-physicist, and in 1993 a reform of the programme was initiated, aimed at enhancing students' problem-solving and communications skills. Such moves are currently common (Baillie, 1999) but in this case it was followed by a drastic drop-out over the two following years, which is the immediate reason for the current study.

The aim of the overall study is to illuminate the factors surrounding the drop-out that followed the reform. It was observed that the existing programme had been compressed in order to accommodate the extra curriculum. A survey of students indicated a puzzling split: that about one third found it to be a stimulating programme and the same number found it grinding. We set out to examine the idea that the large number of courses that go to make up the first year, taught by relatively isolated teachers, led to a fragmentation of the content for some of the students. In particular, we hypothesised that the programme was not experienced as a whole and that becoming a physicist had been relegated to the background, while coping with many disparate courses – predominantly mathematics – had come to the fore.

A more pragmatic aim of the study was to increase the programme board's awareness of the complexity of the first year from the students' perspective, and of the results of their decisions on the conditions for learning.

The research questions addressed by the study as a whole are

- How do students make sense of Physics from the point of view of constituting physics knowledge?

---

[1] By "Physics" with a capital P we denote the programme of Engineering Physics at Chalmers; by "physics" with a small p we mean the physical world and the world of physicists.

- What factors in the programme and the student experience of the programme can be related to the students' approaches to studying and learning?
- What implications are there for the individual student, in their quest for making sense of Physics?

A wider aim is to illuminate the oft expressed goal of educators in designing new programmes or reforming old ones, that they want to improve the integration of the students' knowledge, in particular mathematics and its applications. We ask ourselves what the implementations of this goal might look like from the students' points of view.

This paper focuses on the first of these questions, and the implications for faculty.

## *2. Method*

The study was carried out with a predominantly phenomenographic approach (Marton, 1981; Marton & Booth, 1997). This implies that we were interested in variation in the ways in which the students experienced their first year of study with respect to its content and structure, it being made up of some twelve distinct courses distributed across mathematics, physics and engineering subjects in rough ratio 2:1:1.

Data was collected primarily through interviews. These were held with 20 students sometime in the second year of studies, selected to represent a cross-section of success among those who remained on the programme (and thus including some who had reached the verge of failure). The interview was directed towards exploring the variation of ways in which the students had experienced their first year, with learning the subjects both as individual courses and as an integrated whole. First it probed specifically into the ways in which the student saw the relations between the courses, both structurally and meaningfully, and engaged them in a semi-structured discussion of factors surrounding their studies. The students were initially given an A3 sheeet of paper with all the course titles spread over it, and they were asked to join them up according to perceived relationship. The interviewer began by asking the students for the meaning of the lines they had drawn, giving them opportunity to discuss the relationships between the course contents freely. The interviews then continued by developing the relations they referred to and by taking up students' overall experience of the programme, their reasons for choosing Physics, what kept them going, the way in which the whole was seen, and their approaches to studying.

In line with the phenomenographic approach, the interviews are seen as forming a "pool of meaning" in which the variation in ways of experiencing the phenomena of interest are to be seen. By reading the interviews repeatedly, now as expressions of individual students, now as series of extracts related to specific issues, we delved more and more deeply into the meaning of "Physics" as seen by the students. Categories were formed and reformed; extracts from interviews were sought to support and give substance to the categories; and logical and empirical links between categories were explored. The aim was to offer a hierarchy of empirically grounded and logically consistent categories of description which capture the essence of the whole experience and reveal the essential variational structure of that experience.

## *3. Results*

Throughout the interviews we took up aspects of ways in which the students experienced the objects of their studies – the content of courses, the relations between them, the relation between them and the future (studies and work), understanding and difficulties. We have also taken note of the ways in which students relate to knowledge, to others and to self, in a system which we identify with an ethical aspect of experience.

From the data we have analysed an outcome space of six categories of description, forming a hierarchy of increasing sense-making. We introduce the term "knowledge fragment" to indicate the way in which the students seem to experience what constitutes the courses. These we see being experienced as self-contained pieces, bearing meaning only in a local sense, neither perceived as legitimate or recognisable outside the immediate educational locality. We also use the term "knowledge object", related only partially to the notion as used by Entwistle & Marton (1994). While they mean the ideal visualisable whole "made up of a tightly integrated and structured interconnected ideas and data which together make up our own personal understandings" attained after a good deal of intensive study, we refer rather to the whole that students are experiencing, whatever its character might be. Based on our interpretation of the empirical data we even draw a distinction between "study knowledge object" and "physics knowledge object", where in the first focus is on the process of study, (we could say Physics), the latter is focused on the meaning of study related to the physical world – a figure-ground relationship.

There follows a short description of the six ways of experiencing the first year in terms of the study content, with example extracts from the interviews. Note that we are not categorising individual students, but are analysing the whole experience as it is told by the set of students and illustrating by individual statements.

### *3.1. A. Courses are identified with the study situation*

Here the engineering physics programme has been experienced as a discrete set of courses, a means to the end of a degree. These are related to authority, i.e. teachers and tradition, and common features, such as the ways in which courses were organised.

S9[2] indicates such a way of experiencing the first year, here, for example, relating Mechanics and Strength of Materials:

> S9   It was about moments and suchlike, what can I say, forces here and there, but maybe there wasn't such a big link, they share the subject a bit, yes I suppose so

Later in the same interview the same student refers thus to Complex Analysis and Fourier Analysis, bringing out the common organisational feature:

> S9   Well, what can I say, maybe there's not such an enormous connection, but it feels as if they are the same, more that it is the same sort of organisation in the course, sort of, more than what they are about. I really like those courses, no problems to hand in, no bonus points to chase after, but it's just a case of learning really, and being able to work the problems out.
> I    Is it the teaching more than anything?
> S9   Yes, more than the content maybe. But well, I liked them a lot.

We would dub this a "study knowledge object", but it is taken for granted rather than being a focal concern.

### *3.2. B. One course is seen as a prerequisite for another course*

Courses are now related to their content to the extent that a preordained, correct sequence of acquisition of knowledge fragments is assumed. A "red thread"[3] is sought in terms of needs and demands. Authority for the thread – content and structure – is still the domain of teachers and tradition.

---

[2] The 20 students who were interviewed are identified S1 to S20. I refers to the interviewer

> S7    First Algebra and the maths courses, you can't take them away, just like RealB[4], I didn't enjoy that but it still has to be there anyway. Then the rest of the courses, I think they have to be there, but whether or not you could change their order I don't know, maybe you could. I don't know how important EM is, if you could put it earlier or later, it seems to be important because we've done so much of it so far

Asked if he could see a whole in the set of courses he had taken, S7 refers to the lack of insight into the teacher's intentions:

> S7    Yes, sort of… I can see how they've tried to build it up but I don't know if I see the aim of it, to be honest, I feel I'm moving forward but I don't really know where I'm trying to get to

In discussing courses that had common content S6 refers to perceived shortcomings in authority:

> S6    … but it sometimes feels as though the teachers don't really know what we know and don't know. Have you done this before? And have you taken that up? Like in ENets for example, where they took up Laplace transformations, and they came up an awful lot in the exam, on the first exam in any case, Laplace transformations, but it hadn't, in Complex they hadn't had time to take it up properly, and then in ENets there was no time to do it properly either, there they assumed we had gone through it in Complex. That sort of thing. It's a bit as if, things run into one another a bit vaguely, the boundaries are unclear, between Control and ENets, for example

Compared with the previous category, the emerging "knowledge object" is related to study, now being focused on in trying to find the fit of the fragments.

### *3.3. C. One course is seen as being useful in other courses*

Courses now support one another, but they still are necessarily arranged in a specific order. Reference is made to the knowledge fragments that constitute the courses, which mesh into one another, course-to-course.

> I    And the line between Mechanics and MatStrength?
> S3    It's more a question of MatStrength having a bit of Mechanics in it, … the course in Strength of Materials, I thought that went smoothly, I didn't get any links to any other subjects at all except to just Mechanics, it was mostly force, forces and other things of course, but…

A number of different knowledge fragments, not necessarily from the same course, may build a specific technique or application, and the future usage of such applications comes into the picture.

> S12    I think that whatever courses you choose you can never cover so much that it'll be exactly what you finally work with, there are little bits in each course you have use of and recognise. I don't think it's the details, as long as you're not going to do research, and I have no idea if I'll do that.

Now the knowledge object starts to have features relating to physics, while study as the object still dominates.

---

[3] A "red thread" is a Swedish term for the logical structure that is either planned or apparent. It is a very common term among both students, who demand them, and teachers, who try to make them apparent, both in individual courses and in programmes of courses.

[4] Courses are referred to by abbreviated names in interviews

| Analysis of single variable | RealA | Electro-magnetic fields | EMFields |
|---|---|---|---|
| Analysis of several variables | RealB | Passive & active electric networks | ENets |
| Complex analysis | Complex | Mechanics | Mechanics |
| Vector analysis | Vector | Strength of materials | MatStrength |
| Fourier analysis | Fourier | Automatic control | Control |
| Linear algebra | Linear | Numerical analysis | Numeric |

### *3.4. D. Courses are related through mutual illumination*

Here is to be found sense-making for the first time. Courses now lend meaning to each other and understanding in an earlier course can be found in a later course. There are now networks that mesh and unmesh, knowledge fragments might be grouped together in different ways and offer different perspectives. There is a dynamic in what is focal or non-focal, and thematic or non-thematic. The Physics that is constituted takes on a dynamic form and begins to resemble a "physics knowledge object" rather than a "study knowledge object".

What is met in one course can illuminate or explain what is met elsewhere:

> S12  I see that [ENets] more as a lot of things you just have to accept, currents that go here and there in ENets, they get explained in EMFields. That's what I think is essential when you do the Physics programme, that you get these explanations and don't simply apply things, but you go a bit further

When discussing sudden insights he had had, S3 says:

> S3  Yes, in Numeric as well, when you studied optimisation and other things that you could sort of deduce from the theory from algebra and linear spaces and things, that you could…, there it comes in, you saw that it was that you were working at without thinking of it, and that you'd done it before in RealB as well, without knowing that you were projecting it on a subspace sort of? There I felt sort of Wow, when I did Linear anyway.

S3 refers to his need to put abstractions into context in order to find "physical meaning":

> S3  The relation between Vector and EMFields was really good. I failed the exam when I took it then, in the last quarter, largely because I didn't feel any, sort of had no connection to what it's used for actually. We did take EMFields at the same time, but we didn't get so far that you could start to look around… there was sort of no… you learn a bit about vorticity and so on but it has no physical meaning before you've done the EMFields course. But then when we had learned electro-magnetic theory, learned a whole load of Vector, then the parts of that course started to come together

Being able to confirm abstract concepts in a practical context is referred to:

> S12  There (Electrical Measurements) we measure, in some of the labs, things we learned about in EMFields, phenomena with reflections and suchlike, and see that they do in fact exist, that's a sort of link maybe

### *3.5. E. Courses fit together into an adaptable whole*

The courses are seen as constituting parts of a whole, and the strict ordering structure of the educational programme knowledge content is broken apart. An internal dynamic enables a picture to develop which is different on different occasions, depending on what aspects are brought into focus.

S1, speaking of courses where he has gained understanding:

> S1  It was sort of, you discovered that in some way, like in RealB, that you suddenly can simply transform a two-dimensional [double-]integral to a three-dimensional [triple-]integral at once. Now it feels much more obvious that it is so. It didn't then. To be able to see something in a different way, that you couldn't see before

S10 describes with pleasure tying things together, achieving a "knowledge object" in Marton & Entwistle's sense:

> S10  Well, when we did Complex, and got towards the end of it, you sort of began to see how a lot of it is related to what you studied in the first year, then, you sort of got to tie in lots of the maths courses you'd taken earlier, you got a bird's eye perspective over the whole thing [as you came to the end] of Complex, so you started to feel now, now I see some sort of connections anyway. That was really cool.

And S1 takes s further step in realising that what has been encapsulated in one course can be seen as a special case of a more general field of knowledge; the knowledge object is not only visualisable but reformable when needed:

> S1   It's quite a lot of application. In Control you draw upon examples from Mechanics when you are working out your systems. And in MatStrength it's actually a question of, you actually take your mechanics systems and make them very very small, so that they can't shear and bend. You're taking Mechanics into a new dimension, that's why you use deformable bodies there [in Strength of Materials] instead. Large bodies. That sounded good!

### *3.6. F. Courses in Physics come into physics*

The borders between courses are erased, a physics knowledge object is constituted, physics and the physics world are one with the knower.

> S12   I think you get a lot of ahah-sensations in the EMFields course, you get to understand a lot of things that before you simply accepted. It's really courses like that that are fun to take, you understand how a microwave oven works and suchlike

What is met in courses is related to potential others in potential situations outside university

> S10   That's how it was in Control. There you had to tackle problems and sort of feel that, if we had a specific problem here, something technical that an engineer could come across, how would I solve it? And how good would my solution be? There really ought to be a lot of that, things that an employer wants. You should be able to come up with a solution and then judge your solution critically, and see if it is acceptable. That feels right somehow.

### *3.7. Summary of the provenance of the categories*

Categories arise from the pool of meaning provided by the set of interviews, and not from individual students. We can see, however, extracts from individual interviews that indicate one category or the other. In table 1 we summarise the provenance of the categories.

| | |
|---|---|
| **A** | S2, S9, S18 |
| **B** | S2, S5, S6, S7, S8, S9, S10, S11, S15, S17, S18, S19 |
| **C** | S1, S3, S4, S7, S10, S11, S12, S13, S14, S16, S17, S20 |
| **D** | S1, S3, S4, S8, S11, S12, S13, S14, S16, S17, S20 |
| **E** | S1, S10, S12, S16 |
| **F** | S10, S12, S16 |

Table 1. Individual interviews indicate a range of categories

### *4. Discussion*

The empirical study has resulted in an six-tiered outcome space of ways in which students of Physics experience their first year of study, which is a hierarchy of sense-making. The first three (A, B, C) refer to courses as courses, knowledge fragments being the components of the courses, isolated in A, building on one another in B, and meshing into one another in C. The second group of three (D, E, F) bring the meaning of the content into focus and ascribe different relationships between the content – mutual in D, multiple in E and finally extending outside the programme to physics phenomena in F. The similarity to studies of conceptions of learning is striking (Marton et al., 1993, Säljö, 1979), in that meaning, or sense-making, is a watershed between two groups of three categories.

We have introduced the notions of "study knowledge object" and "physics knowledge object" to distinguish between making sense of the study situation in one way or another, and making sense of physics. To varying degrees these two aspects of the knowledge object are present throughout the categories, but "study" dominates the earlier categories and "physics" becomes increasingly in focus in the latter categories.

### *4.1. The structure of the experience of the first year of Physics*

In Table 2 we have analysed the results according to the structure of experience (Marton & Booth, 1997). It is seen that the referential aspect indicates clearly the shift from no-meaning to meaning between C and D. The external horizon of the structural aspect of the ways of experiencing shifts gradually from an unproblematised studying at the university, here and now, through a refocusing on future study and the world outside the university, to finally embrace physics as a way of seeing the world outside the university. The internal horizon of the structural aspect – how the parts of the ways of experiencing are related to one another and to the whole – shifts in a more discrete sense. Isolated, or possibly grouped, fragments are all there are in A, the blocks taking on a linear preordained arrangement in B. In C, thanks to overlapping fragments, the preordained linear arrangement has branches and parallel paths as well, while in D the fragments are related more by meshing facilitated by understanding, thus giving freedom for realignment and restructuring. In E forms of knowledge are constituted of the fragments to be found in courses, which give new perspectives and ways of seeing, while in F these ways of seeing are directed outside current experience to an unknown future.

Based on the analytical device of the phenomenographic structure of experience, we have extended the analysis to consider the nature of knowledge, drawing largely on the characteristics of the internal horizon of the structure of the ways of experiencing the first year of Physics. Further, we consider, following Perry's seminal work "Epistemological and ethical development in the college years" (Perry, 1970/99) an ethical aspect of the experience, drawing largely on the referential aspect.

Let us relate the categories to the individuals who were interviewed. If we look back to Table 1 we see that almost all students expressed experiencing the first year in more than one of these ways, and most expressed ways that fall above and below the "sense-making" watershed. That so many voiced C, even if mainly speaking of sense-making, is hardly surprising giving the design of the interview, based as it was on a chart of individual courses. Of the 20 interviewed, 8 students expressed ways of experiencing their first year of Physics only in the range A to C, which can be interpreted as their not being competent to see the first year in a sense-making way. What these also have in common is reference to the weight of studies and the effects it has had on them. S2, an ambitious student not content to get less than top grades and having chosen Physics because it is reputed to be the toughest programme, says at the end of his interview:

> S2   Sometimes it feels as though there's much too much to do. You can understand that a lot drop out. And there are periods when you can never take time off, there're always things to do but you don't have time. Then it is easy to lose interest and go over to something else instead … when you get to exams you generally have to learn what you need to and it often feels that during the study quarters you are mostly behind and don't know anything.

S6, less confident of her abilities relative to her peers, says:

> S6   Interest has been killed by the tempo.

|   | Structural Aspect | | Referential aspect | Nature of knowledge | Ethical aspect |
|---|---|---|---|---|---|
|   | *External horizon* | *Internal horizon* | | | |
| **A** | University | Courses, tasks, organisation, teachers, exams, | Gaining a degree | Isolated fragments, encapsulated in courses | Authority with teachers. "We need the degree" |
| **B** | University, future years of Physics | Courses, red threads, | Building up the programme according to the teachers' intentions | Ordered fragments | Authority with teachers. "Knowledge is what they want us to find" |
| **C** | University, future years of Physics, world of work | Courses, red threads, overlap and application | Building up the programme according to the teachers' intentions | Fitting fragments | Authority with teachers "Knowledge is there to be put together" |
| **D** | University, future years of Physics, world of work | Knowledge fragments, related by explanation theoretical reasoning and confirmed by empirical evidence | Gaining an understanding of the basics of the programme | Meshed and re-arrangeable fragments integrated by understanding | Responsibility shift towards self. "Knowledge is there to understand" |
| **E** | University, future years of Physics, world of work | Knowledge forms that give ways of seeing | Gaining new ways of seeing | Knowledge object forming | Responsibility with self. "Knowledge is ways of seeing" |
| **F** | University, future years of Physics, physics phenomena, world of work | Knowledge forms that give ways of seeing physics | Gaining physics ways of seeing | Knowledge object related to self and the physics world | Commitment to physics a possibility. "Knowledge is a way of experiencing the world" |

Table 2. Analysis of the variation in ways of experiencing the first year of Physics, with respect to learning physics

One extension to this work has to be to make contact with students who have actually dropped out and see how their ways of experience fit into and extend this picture. Another is to look at the results in case studies of individual students.

### *4.2. Ethical aspect of the experience of the first year of Physics*

The clear watershed between category C and D is further emphasised if an ethical aspect of the categories is taken into account. The different interpretations of "authority" implies different views of knowledge. By "authority" we mean where the responsibility lies for the structure and outcome of the first year of study. In the first group of categories (A,B,C) the authority clearly lies outside of the student, the responsibility and problem formulation privilege are mainly taken by the teachers and other persons "in power", not necessarily known to the student. Following their guidance, the student is guaranteed a successful outcome of the studies. In the second group of categories (D,E,F), the responsibility is taken and agenda is set mainly by the student. Drawing upon the work by Perry (1999), this is very similar to his developmental scheme from the dualistic world of Authorities and Absolutes to the relative world of Commitment and Nuances.

Parallel to the responsibility aspect, different "coping strategies" could be observed. Even though the same physical act might exist in both groups, e.g. solving old exams (with given solutions) close before the exam (popularly called "tentakit"~"examfix"), the context is very different. In the first group this is one of the acts done to guess what "they", i.e. the authorities, teachers, want, but in the second this is a opportunity to delve into more complex problems with a context possibly easier to relate to earlier knowledge. We see these strategies as ways of creating a confidence, an assurance, trying to take control over the situation as it is perceived and bring a sense of purpose to one's studies.

This leads us to relate the dichotomous approach to study – deep approaches vs. surface approaches (Marton et al., 1984) – to the individual's perception of authority and the source of the sense of control and/or self-assurance. A student who perceives authority for knowledge lying outside himself will seek ways of satisfying that authority – finding the "red thread" that teachers have built their courses round, trying to build knowledge fragments into a coherent whole according to *their* plan by studying *their* exam solutions, by reading over and over *their* notes and text-books – a classic surface approach in which attention is paid to the tokens. A student who sees the authority lying partly at least with himself will focus on the meaning of and relationships between knowledge fragments using strategies of studying exam solutions to see the variation in ways the fragments can mesh to one another, reading notes and text-books to spy hitherto unremarked connections – the classic deep approach of seeking what the tokens signify.

We intend to extend this research with a study specifically aimed at studying the ethical aspects of students' study, their experience of authority and ways of coping with the need for assurance.

### *4.4. Conditions for learning and implications for faculty*

The Physics programme is the major factor in creating the conditions for learning for these students. The curriculum, embodied as it is in courses and teaching, is the major contributor to the students learning physics, becoming engineering physicists in knowledge, language and culture. While this study is not able to say much about individual courses and individual teachers, and their effects on the conditions for learning, one can conclude from it that the programme as a whole, and how it is organised and conducted, has a profound effect.

Any programme that is organised as this one is, as a set of courses given by subject specialists, (and degree programmes mostly are) has to have as an overriding goal that the students come to see the subject matter as a related whole, and that this provides them with ways of seeing and coping with an as yet unknown world. This issue has been argued cogently by Bowden and Marton (1998)

This study has a clear aim, which is to lead to improvements in the study situation for Physics students by informing and influencing the teachers and the leaders of the programme. The vision of the programme is to produce all-round engineering physicists, capable of working in a wide field of engineering research, development and leadership. The goals of the programme are less clearly articulated. An oft-stated desire of programme leaders, not least Physics, is to encourage an integration of knowledge so that students come to an understanding of a whole from the parts that are presented in individual courses, yet neither the goals of the programme nor the goals of individual courses take this line. And, as we see from this study, the desired integration is not a self-evident result, even when courses are arranged to offer different aspects of a particular phenomenon.

A naive belief in a given structure, known to and enforced by external authority, works against integration on the large scale, as it works against deep approaches on the small scale. There are examples of groups of teachers who try hard to build "red threads" into their programmes, but fail to ask "whose red thread?" The evidence from this study shows that a red thread can be experienced as a security line to be clung to rather than an integration guide through the constituents of an emerging physics knowledge object. Where integration becomes possible is in Category D, when knowledge fragments are perceived to mesh and unmesh like Lego blocks, as appropriate for current purpose.

The main aim as we see it should be to encourage and support the students to develop a commitment to Physics and physics. How, though, can teachers create a study situation that disfavours the early categories with their strategies of coping in order to bring disjointed bits of knowledge into the pattern demanded by external authority, and favours later categories in which there is a commitment to understanding and making a coherent adaptable whole of the fragments through which new phenomena can be seen and integrated to form a new whole?

The least but first step is to create a new, and hitherto lacking, College of teachers which goes across department boundaries, and where the whole programme and integration from the students' perspective is the theme. This is in line with the recommendations of Bowden & Marton (1998) who propose academic teams for curriculum design, cemented by a system of quality assurance that gives both team and individual responsibilities. That the teachers learn about one another's subjects and – above all – about their students' learning, and to relate this to a theoretical framework for learning, needs to be the goal of the new forum. If we see this in terms of knowledge objects, we can say that the teachers are thus engaged in building a knowledge object which they will offer to their students.

Another, more focused, approach we can refer to is that offered by Alant et al. (Alant et al., 1999) Focusing on the observed tendency for students to spend time on quantitative, algorithmic aspects of their physics studies at the expense of exploring the qualitative aspects that lead to an understanding, they devised and studied the results of a teaching experiment. The teachers on the course adopted strategies that would foster a conceptual focus, foster reflection on the nature of the discipline, promote reflection on the value and relevance of what was taught, foster student activity, and make metacognition explicit. The results showed

> that the students' approaches to learning physics and their conceptions of learning shifted dramatically away from the rote-memorising with which they generally entered higher education. In addition, linked to shifts in the students' conceptions of learning were shifts in the ways in which they conceived of the nature of science, moving from an 'immutable' conception of science to a more 'tentative' conception of science

They also relate their results to Perry's work (Perry, 1970/99), pointing out the epistemological variation that they observed. A distinct difference between the study of Alant et al. and the present study is that whereas they were looking at students taking close-knit single and extensive physics courses, the students we were looking at were meeting a large number of teachers in a large number of separate courses. The content of their strategies, however, could well make a substantial contribution to the activities of a suggested College of teachers.

## *Conclusion*

The question we addressed was "How do students make sense of Physics from the point of view of constituting physics knowledge?". We have identified six qualitatively different ways in which students experience the first year of Physics and analysed the variation in terms of the structure of experience, the nature of knowledge and an ethical aspect related to the identification of authority. Three of these ways of experiencing the first year are considered to be unproductive in terms of making sense of physics, while the other three support to an increasing degree the formation of a well-grounded physics knowledge object, where fragments from different courses are integrated through ways of seeing physics. The ethical aspects have potentially profound implications for the ways students take on their studies, related in some sense to the deep and surface approach dichotomy, and deserve further investigation.

## *6. References*

Alant, B., Linder, C., & Marshall, D. (1999). *Metacognitive-linked developments arising from the design and teaching of conceptual physics*. Paper presented at 8th European Conference for Research on Learning and Instruction, August 24-28, 1999, Göteborg, Sweden

Baillie, C. (1998). Addressing first year issues in engineering education. *European Journal of Engineering Education*, 23, 4, 453-465

Bowden, J. & Marton, F. (1998). *The University of Learning. Beyond quality and competence in higher education.* London: Kogan Page.

Entwistle, N. & Marton, F. (1994). Knowledge objects: understandings constituted through intense academic study. *British Journal of Educational Psychology, 64*, 161-178.

Marton, F: (1981) Phenomenography – describing conceptions of the world around us. *Instructional Science*, 10, 177-200

Marton, F., Beaty, E. & Dall'Alba G. (1993). Conceptions of learning. *International Journal of Educational Research*, 19, 277-300.

Marton, F. & Booth, S. (1997). *Learning and Awareness*. Mahwah: Lawrence Erlbaum Ass.

Marton, F., Hounsell, D. & Entwistle, N. (Eds.) (1984). *The Experience of Learning*. Edinburgh: Scottish Academic Press.

Perry, W. (1970/99). *Forms of ethical and intellectual development in the college years.* San Francisco: Jossey-Bass Inc.

Säljö, R. (1979). Learning in the learner's perspective. I. Some common-sense conceptions. *Reports from the Department of Education, Göteborg University, No 76.*